%% file: pcs_tsp_final.tex
\documentclass[10pt,journal,pdftex]{IEEEtran}
\usepackage{cite}

\input{myMacros}

\def\integers{\mathbb{Z}}
\def\cA{{\cal A}}

\def\dotprod#1#2{\langle #1, #2 \rangle}

\begin{document}

\title{Compressed Sensing Performance Bounds\\
Under Poisson Noise}
\author{Maxim Raginsky,~\IEEEmembership{Member,~IEEE},
Rebecca M.~Willett,~\IEEEmembership{Member,~IEEE},\\
Zachary T.~Harmany,~\IEEEmembership{Student Member,~IEEE}, and
Roummel F.~Marcia,~\IEEEmembership{Member,~IEEE}
\thanks{This work was supported by NSF CAREER Award No. CCF-06-43947, DARPA
Grant No. HR0011-07-1-003, and NSF Grant DMS-08-11062. Portions of this work were presented at the IEEE International Symposium on Information Theory, Seoul, Korea, June/July 2009, and at the IEEE Workshop on Statistical Signal Processing, Cardiff, Wales, UK, August/September 2009.}
\thanks{M.~Raginsky, R.M.~Willett, and Z.T.~Harmany are with the Department of Electrical and Computer Engineering, Duke University, Durham, NC 27708 USA (e-mail: m.raginsky@duke.edu, willett@duke.edu, zth@duke.edu).}
\thanks{R.F.~Marcia is with the School of Natural Sciences, University of California, Merced, CA 95343 USA (e-mail: rmarcia@ucmerced.edu).}
}%

\maketitle

\begin{abstract}
  This paper describes performance bounds for compressed sensing (CS)
  where the underlying sparse or compressible (sparsely approximable) signal is a vector of nonnegative intensities whose
  measurements are corrupted by Poisson noise. 
  In this setting, standard CS techniques cannot be applied directly for several
  reasons.  First, the usual signal-independent and/or bounded noise models
 do not apply to Poisson noise, which is
  non-additive and signal-dependent.  Second, the CS matrices
  typically considered are not feasible in real optical systems
  because they do not adhere to important constraints, such as nonnegativity and photon
  flux preservation.  Third, the typical
  $\ell_2$--$\ell_1$ minimization leads to overfitting in the
  high-intensity regions and oversmoothing in the low-intensity areas.
  In this paper, we describe how a feasible positivity- and flux-preserving
  sensing matrix can be constructed, and then analyze the performance
  of a CS reconstruction approach for Poisson data
  that minimizes an objective function consisting of a negative
  Poisson log likelihood term and a penalty term which 
  measures signal sparsity.  We show that, as the overall
  intensity of the underlying signal increases, an upper bound on the
  reconstruction error decays at an appropriate rate
  (depending on the compressibility of the signal), but that for a
  fixed signal intensity, the signal-dependent part of the error bound actually
  grows with the number of measurements or sensors. This surprising
  fact is both proved theoretically and justified
  based on physical intuition.\\
  
    

\end{abstract}

\thispagestyle{empty}

\section{Introduction}
\label{sec:intro}

The basic idea of compressed sensing (CS) is that, when the signal of
interest is very sparse (i.e.,\ zero-valued at most locations) or
highly compressible in some basis, relatively few ``incoherent''
observations are sufficient to reconstruct the most significant
non-zero signal components \cite{CS:candes2,CS:donoho}. Despite the
promise of this theory for many applications, very little is known
about its applicability to photon-limited imaging systems, where
high-quality photomultiplier tubes are expensive and physically large,
limiting the number of observations that can reasonably be acquired by
the system.  Limited photon counts arise in a wide variety of
applications, including infrared imaging, nuclear medicine, astronomy
and night vision, where the number of photons detected is very small
relative to the number of pixels, voxels, or other entities to be
estimated. Computational optics techniques, compressed sensing
principles, and robust reconstruction methods can potentially lead to
many novel imaging systems designed to make the best possible use of
the small number of detected photons while reducing the size and cost
of the detector array.  Recent work has {\em empirically} explored CS
in the context of photon limited measurements
\cite{fesslerPCS,harmanyPCS,marioPCS,starckPCS}, but theoretical
performance bounds similar to those widely cited in the conventional
CS context previously remained elusive.

This is in part because the standard assumption of signal-independent
and/or bounded noise (cf. \cite{CS:noiseEC,CS:noiseRN}) is violated
under the Poisson noise models used to describe images acquired by
photon-counting devices \cite{SHW93}. The Poisson observation model
\begin{equation}
y \sim \Poisson(A\tf), \label{eq:Poisson2}
\end{equation}
where $\tf \in \reals^{m}_+$ is the signal or image of interest, $A \in
\reals^{N \times m}$ linearly projects the scene onto an
$N$-dimensional space of observations, and $y \in \integers^N_+$
is a length-$N$ vector of observed Poisson counts, stipulates that the likelihood of observing a
particular vector of counts $y$ is given by
$$
p(y|A\tf) = \prod^N_{j=1} \frac{(A\tf)_j^{y_j}}{y_j!} e^{-(A\tf)_j},
$$
where $(A\tf)_j$ is the \jth component of $A\tf$. Moreover, in order to correspond to a physically realizable linear optical system, the measurement matrix $A$ must be:
\begin{itemize}
	\item {\bf Positivity-preserving} --- for any nonnegative input signal $f$, the projected signal $Af$ must also be nonnegative. Using the standard notation $f \succeq 0$ to denote the nonnegativity of $f$, we can write this condition as
	$$
	f \succeq 0 \qquad \Longrightarrow \qquad Af \succeq 0.
	$$
      \item {\bf Flux-preserving} --- for any input signal $f \succeq
        0$, the mean total intensity of the observed signal $Af$ must
        not exceed the total intensity incident upon the system:
	$$
	\sum^N_{i=1} (Af)_i \le \sum^m_{i=1} f_i.
	$$
\end{itemize}

\subsection{Surprising main result}
\label{ssec:surprise}

In this paper, we make the following contributions:
\begin{itemize}
\item design physically realizable sensing matrices, $A$, which
  incorporate the above positivity and photon flux
  preservation constraints;
\item propose a penalized-likelihood objective function for
  reconstructing $\tf$ from $y$ observed according to (\ref{eq:Poisson2});
\item derive upper bounds on the error between $\tf$ and the estimate
  $\hf$ and demonstrate how the error scales with the overall intensity ($I
  \deq \sum_i \tf_i$), the size of $\tf$ ($m$), the number of
  measurements ($N$), and the compressibility of the signal in some
  basis ($\alpha$); and
\item present empirical results demonstrating the efficacy of the
  proposed method.
\end{itemize}
In particular, the main theoretical result presented in this paper
shows that, for an $\alpha$-compressible signal of total intensity
$I$\footnote{\rmw{More precisely, $I$ refers to the total intensity
    integrated over the exposure time, so that increasing $I$ can be
    associated with more source intensity, longer exposure time per
    measurement, or both.}},
$$\mbox{reconstruction error} \propto N \left(\frac{\log
    m}{I}\right)^{\frac{2\alpha}{2\alpha+1}} + \frac{\log(m/N)}{N} $$
for $N$ sufficiently large. (As we show in Section~\ref{sec:rates},
there is a threshold effect in that the number of measurements $N$
must be large enough to guarantee that the {\em per-sensor}
reconstruction error decays with the incident signal intensity $I$.)
Since the total number of observed events or photons, $n \deq
\sum_{i=1}^N y_i$, is the realization of a Poisson random variable
with intensity $I$, the bound reflects how error scales with the
number of observed events.

While the rate of error decay as a function of the total intensity,
$I$, coincides with earlier results in denoising contexts, the
proportionality of the intensity-dependent term in the error to $N$
may seem surprising at first glance. However, one can intuitively
understand this result from the following perspective. If we increase
the number of measurements ($N$) while keeping the expected number of
observed photons ($I$) constant\footnote{\rmw{In some systems, such as
    a single-detector system, more measurements might seem to suggest
    more observed photons. However, holding $I$ fixed while increasing
    $N$ implies that each measurement is collected over a shorter
    exposure. Thus increasing $N$ does {\em not} correspond to an
    increase in the number of observed events/photons.}}, the number
of photons per sensor will decrease, so the signal-to-noise ratio
(SNR) at each sensor will likewise decrease, thereby degrading
performance. Having the number of sensors exceed the number of
observed photons is not necessarily detrimental in a {\em denoising}
or {\em direct measurement} setting (i.e., where $A$ is the identity
matrix) because multiscale algorithms can adaptively bin the noisy
measurements together in homogeneous regions to achieve higher SNR
overall \cite{willett:riskb,willett:tmi03}. 
However, in the CS setting the signal is first altered by the
compressive projections in the sensing matrix $A$, and the raw
measurements cannot themselves be binned to improve SNR. In
particular, there is no natural way to aggregate measurements across
multiple sensors because the aggregation effectively changes the
sensing matrix in a way that does not preserve critical properties of
$A$.

One might also be surprised by this main result because in the case
where the number of observed photons is very large (so that SNR is
quite high and not a limiting factor), our bounds do not converge to
the standard performance bounds in CS. This is because our bounds
pertain to a sensing matrix $A$ which, unlike conventional CS matrices
based on i.i.d.\ realizations of a zero-mean random variable, is
designed to correspond to a feasible physical system. In particular,
every element of $A$ must be nonnegative and appropriately scaled, so
that the observed photon intensity is no greater than the photon
intensity incident on the system (i.e., we cannot measure more light
than is available). This rescaling dramatically impacts important
elements of any performance bounds, including the form of the
restricted isometry property \cite{RIP,JustRelax}, even in the case of
Gaussian or bounded noise. (Additional details and interpretation are
provided in Section~\ref{sec:dc} after we introduce necessary concepts
and notation.)

As a result, incorporating these real-world constraints into
our measurement model has a {\em significant and adverse impact on the
  expected performance of an optical CS system}.

\subsection{Relation to previous CS performance bounds}
\label{ssec:previous}

The majority of the compressed sensing literature assumes that there
exists a ``sparsifying'' reference basis $W$, so that $\ttheta \deq
W^T \tf$ is sparse or lies in a weak-$\ell_p$ space.  When the matrix
product $AW$ obeys the so-called {\em restricted isometry property}
(RIP) \cite{RIP,JustRelax} or some related criterion, and when the
noise is bounded or Gaussian, then $\ttheta$ can be accurately
estimated from $y$ by solving the following $\ell_2-\ell_1$
optimization problem (or some variant thereof):
\begin{equation}
\widehat{\theta} 
= 
\argmin_{\theta} \left[ \|y-AW\theta\|_2^2 + \tau \|\theta\|_1 \right],
\label{eq:nlp2}
\end{equation}
where $\tau > 0$ is a regularization parameter
\cite{CS:donoho,LASSO,JustRelax}. 

However, the $\ell_2$ data-fitting term, $\|y-AW\theta\|_2^2$, is
problematic in the presence of Poisson noise.  Because under the
Poisson model the variance of the noisy observations is proportional
to the signal intensity, $\ell_2$ data-fitting terms can lead to
significant overfitting in high-intensity regions and oversmoothing in
low-intensity regions.  Furthermore, photon-limited imaging systems
impose hard constraints on the nature of the measurements that can be
collected, such as non-negativity, which are not considered in much of
the existing compressed sensing literature (recent papers of Dai and
Milenkovic \cite{DaiMil09} and of Khajehnejad et al. \cite{KDXH09} are
notable exceptions).  Bunea, Tsybakov and Wegkamp
\cite{wegkampDensity} study the related problem of using $\ell_1$
regularization for probability density estimation, but rather than
assuming incoherent measurements of a random variable (similar to our
setup), they assume direct observations of a random variable and
learn, for example, a sparse mixture model. Furthermore, their work
assumes infinite precision in the observed realizations of the random
variable, so that their analysis does not provide any insight into how
the number or bit depth of detector elements impacts performance.
More recent work by Rish and Grabarnik \cite{expFamCS} explores
methods for CS reconstruction in the presence of exponential family
noise using generalized linear models, but does not account for the
physical constraints (such as flux preservation) associated with a
typical Poisson setup.

In this paper, we propose estimating $\tf$ from $y$ using a
regularized Poisson log-likelihood objective function as an
alternative to (\ref{eq:nlp2}), and we present risk bounds for
recovery of a compressible signal from Poisson
observations. Specifically, in the Poisson noise setting we maximize
the log-likelihood while minimizing a penalty function that, for
instance, could measure the sparsity of $\theta=W^Tf$:
\begin{equation}
\begin{array}{rll}
\displaystyle
	\hf \ = \ & \displaystyle \argmin_{f}  & \displaystyle
	 \sum_{j=1}^{N} \left [(Af)_j-y_j \log (Af)_j \right ] + \tau \pen( f ) \\
	& \textrm{subject to} &  
	f \succeq 0, \;  \sum_{i=1}^m f_i = I
\end{array} \label{eq:CSP2} 
\end{equation}
where $\pen(\cdot)$ is a penalty function that will be detailed later,
and $I$ is the known total intensity of the unknown $\tf$.

\subsection{Organization of the paper}
\label{ssec:org_paper}

\rfm{Section~\ref{sec:problem} contains the problem formulation, describes the proposed approach, and details the construction and properties of a feasible sensing matrix $A$. In Section~\ref{sec:risk_bounds} we develop an
oracle inequality for the proposed estimator and then use it to
establish risk bounds for compressible signals. Section~\ref{sec:experiments} contains a
proof-of-concept experiment based on recent breakthroughs in sparse
reconstruction methods we initially proposed in \cite{harmanyPCS}. For the sake of readability, proofs of all theorems are relegated to the appendices.}

\section{Problem formulation and proposed approach}
\label{sec:problem}

We have a signal or image $\tf \succeq 0$ of size $m$ that we wish to
estimate using a detector array of size $N \ll m$. We assume that the total intensity of $\tf$, given by $I \deq \| \tf \|_1 = \sum^m_{i=1} \tf_i$, is known {\em a priori}. We make Poisson observations of $A\tf$, $y \sim \Poisson(A\tf)$, where $A \in \reals^{N \times m}$ is a positivity- and flux-preserving sensing matrix. Our goal is to estimate $\tf \in \reals^m_+$ from $y \in
\integers_+^N$. 

\subsection{Construction and properties of the sensing matrix}
\label{sec:A_prop}

We consider sensing matrices composed of zeros and (scaled) ones,
where $p$ is the probability of having a zero and $1-p$ is the
probability of having a one. In the context of optical systems, small
$p$ corresponds to sensing matrices with many ones, which allow most
of the available light through the system to the
detectors. Conversely, large $p$ corresponds to having each
measurement being the sum of a relatively small number of elements in
the signal of interest. To explore the tradeoff between these two
regimes, we explicitly consider $p$ throughout our analysis.

We construct our sensing matrix $A$ as follows. Given some $p \in (0,1)$, let $\nu_p$ denote the probability distribution of a random variable that takes values
\begin{align*}
	\lambda^-_p \deq -\sqrt{\frac{1-p}{p}} & \text{ with probability } p; \\
	\lambda^+_p \deq \sqrt{\frac{p}{1-p}} & \text{ with probability } 1-p.
\end{align*}
Note that $\nu_{1/2}$ is the usual Rademacher distribution which puts equal mass on $-1$ and on $+1$. 
Let $Z$ be an $N\times m$ matrix whose entries $Z_{i,j}$ are drawn i.i.d.\ from $\nu_p$. We observe that
\begin{align}\label{eq:isotropy}
\expect Z_{i,j} = 0 \qquad \text{and} \qquad \expect Z_{i,j} Z_{k,\ell} = \delta_{ik} \delta_{j\ell}
\end{align}
for all $1 \le i,k \le N$ and $1 \le j,\ell \le m$. Most compressed sensing
approaches would proceed by assuming that we make (potentially noisy)
observations of $\tA \tf$, where $\tA \deq Z/\sqrt{N}$. However, $\tA$ will, with high probability, have at least one negative entry, which will render this observation model physically unrealizable in photon-counting
systems. Therefore, we use $\tA$ to generate a feasible sensing matrix $A$ as follows. Let $\1_{r \times s}$ denote the $r
\times s$ matrix all of whose entries are equal to 1. Then we take
$$
A \deq \sqrt{\frac{p(1-p)}{N}}\tA + \frac{1-p}{N} \1_{N \times m}.
$$
We can immediately deduce the following properties of $A$:
\begin{itemize}
	\item It is positivity-preserving because each of its entries is either $0$ or $1/N$.
	\item It is flux-preserving, i.e., for any $f \in \reals^m_+$ we have
	\begin{equation}
	\|Af\|_1 \leq \|f\|_1.
	\label{eq:flux}
	\end{equation}
	This is easy to see directly: if $f \succeq 0$, then $Af \succeq 0$, and
	\begin{align*}
		\| Af \|_1 = \sum^N_{i=1}\sum^m_{j=1} A_{i,j}f_j \le \sum^m_{j=1} f_j \equiv \| f \|_1.
	\end{align*}
\item With probability at least $1-Np^m$ (w.r.t.\ the realization of $\{Z_{i,j}\}$), every row of $A$ has at
  least one nonzero entry. Assume that this event holds. Let $f \in \reals^m$ be an arbitrary
  vector of intensities satisfying $f \succeq (cI)\1_{m \times 1}$ for
  some $c > 0$. Then
\begin{equation}
  A f \succeq \frac{cI}{N}\1_{N \times 1}.
\label{eq:pos_intens}
\end{equation}
To see this, write
$$
	(Af)_i = \sum^m_{j=1} A_{i,j}f_j \ge \frac{1}{N} \cdot \min_{1 \le j \le m}f_j 
	\ge \frac{cI}{N}.
	$$

\end{itemize}

\noindent Furthermore, and most importantly, with high probability $\tA$ acts
near-isometrically on certain subsets of $\reals^m$. The usual formulation of this phenomenon is known in the compressed
sensing literature as the {\em restricted isometry property} (RIP)
\cite{RIP,JustRelax}, where the subset of interest consists of all
vectors with a given sparsity. However, as shown recently by Mendelson
{\em et al.}~\cite{MenPajTom07,MenPajTom08}, the RIP is, in fact, a
special case of a much broader circle of results concerning the
behavior of random matrices whose entries are drawn from a {\em subgaussian
isotropic ensemble}. These terms are defined in Appendix~\ref{app:subgaussian_proof}, where we also prove the following result:

\begin{thm}\label{thm:subgauss_bernoulli} Consider the 
  matrix $\tA = Z/\sqrt{N}$, where the entries of $Z$ are
  drawn i.i.d.\ from $\nu_p$. Then there exist absolute constants
  $c_1,c_2>0$ such that the bound
\begin{equation}
\| u - v \|^2_2 \le 4 \| \tA u - \tA v \|_2^2  + \frac{2 c_2^2 \zeta^4_p \log (c_2 \zeta^4_p m/N)}{N}
\label{eq:subgauss_1}
\end{equation}
will hold simultaneously for all $u,v \in B^m_1$ with probability at least
$1-e^{-c_1 N/\zeta^4_p}$, where $B^m_1 \deq \{ u \in \reals^m : \| u \|_1 = 1 \}$ and 
\begin{align}
\zeta_p \deq \begin{cases}
\sqrt{\frac{3}{2p(1-p)}}, & \text{if } p \neq 1/2 \\
1, & \text{if } p = 1/2.
\end{cases}
\label{eq:zetap}
\end{align}
Moreover, there exist absolute constants $c_3,c_4>0$ such that for any
finite $T \subset S^{m-1}$, where $S^{m-1} \deq
\{ u \in \reals^m : \| u \|_2 = 1\}$ is the unit sphere $(\ell_2)$,
\begin{equation}
1/2 \le  \| \tA u \|^2_2 \le 3/2, \qquad \forall u \in T
\label{eq:subgauss_2}
\end{equation}
holds with probability at least $1-e^{-c_3 N/\zeta^4_p}$, provided $N
\ge c_4 \zeta^4_p \log |T|$.
\end{thm}

\rmw{
\subsection{DC offset and noise}
\label{sec:dc}
The intensity underlying our Poisson observations can be expressed as
\begin{align*}
A\tf =\sqrt{\frac{p(1-p)}{N}}\tA\tf + \frac{I(1-p)}{N} \1_{N \times
  1}.
\end{align*}
As described in Theorem~\ref{thm:subgauss_bernoulli}, the {\em
  idealized} sensing matrix $\tA$ has a RIP-like property which can
lead to certain performance guarantees if we could measure $\tA\tf$
directly; in this sense, $\tA\tf$ is the {\em informative} component
of each measurement. However, a constant DC offset proportional to $I$
is added to each element of $\tA\tf$ before Poisson measurements are
collected, and elements of $\tA\tf$ will be very small relative to
$I$. Thus the intensity and variance of each measurement will be proportional to
$I$,  overwhelming the informative elements of
$\tA\tf$.\footnote{\rmw{We would like to thank Emmanuel Cand\`es and an anonymous
  reviewer for helpful insights on this point.}} (To see
this, note that $y_i$ can be approximated as $(A\tf)_i +
\sqrt{(A\tf)_i}\xi_i$, where $\xi_i$ is a Gaussian random variable with
variance one.) }

\rmw{As we will show in this paper, the Poisson noise variance
  associated with the DC offset, necessary to model feasible
  measurement systems, leads to very different performance guarantees
  than are typically reported in the CS literature. The necessity of a
  DC offset is certainly not unique to our choice of a Rademacher
  sensing matrix; it has been used in practice for a wide variety of
  linear optical CS architectures
  (cf. \cite{riceCamera,gehm:ddis,coifmanCamera,neifeld2007}).  A
  notable exception to the need for DC offsets is the expander-graph
  approach to generating non-negative sensing matrices, which has
  recently been applied in Poisson CS settings \cite{jafarpour:PCS};
  however, theoretical results there are limited to signals which are
  sparse in the canonical (i.e. Dirac delta or pixel) basis.}

\rmw{  {\em As
    a result, the framework and bounds established in this paper have
    significant and sobering implications for any linear optical CS
    architecture operating in low-light settings. } }
\subsection{Reconstruction approach and bounds}
We propose solving
the following optimization problem:
\begin{equation}
\hf \deq \argmin_{f \in \Gamma} \Big[ -\log p(y|Af) +
  2\pen(f) \Big],
\label{eq:opt}
\end{equation}
where $\pen(f) \ge 0$ is a penalty term.  Here, $\Gamma \equiv
\Gamma(m,I)$ is a countable set of feasible estimators $f \in
\reals^m_+$ satisfying $\sum^m_{i=1} f_i = I$, and the penalty
function satisfies the {\em Kraft inequality}:
\begin{equation}
\sum_{f \in \Gamma} e^{-\pen(f)} \leq 1. \label{eq:kraft}
\end{equation}
\rmw{(In (\ref{eq:nlp2}) and (\ref{eq:CSP2}), $\tau$ is a free parameter that
could be selected by the user, while in (\ref{eq:opt}) it is fixed
at 2 for a penalty function that satisfies the Kraft inequality. In
practice one often prefers to use a value of $\tau$ different from what
is supported in theory because of slack in the bounds.)}  While the
penalty term may be chosen to be smaller for sparser solutions $\theta
= W^T f$, where $W$ is an orthogonal matrix that represents $f$ in its
``sparsifying'' basis, our main result only assumes that
(\ref{eq:kraft}) is satisfied. \rmw{In fact, a variety of penalization
  techniques can be used in this framework; see
  \cite{willett:riskb,kalyani:siam,MDL} for examples and discussions
  relating Kraft-compliant penalties to prefix codes for
  estimators. Many penalization or regularization methods in the
  literature, if scaled appropriately, can be considered prefix
  codelengths.}  We can think of (\ref{eq:opt}) as a
discretized-feasibility version of (\ref{eq:CSP2}), where we optimize
over a countable set of feasible vectors that grows in a controlled
way with signal length $m$.

We will bound the accuracy with which we can estimate $\tf/I$; in
other words, we focus on accurately estimating the {\em distribution}
of intensity in $\tf$ independent of any scaling factor proportional
to the total intensity of the scene, \rmw{which is typically of
  primary importance to practitioners. Since the total number of
  observed events, $n$, obeys a Poisson distribution with mean $I$,
  estimating $I$ by $n$ is the strategy employed by most
  methods. However, the variance of this estimate is $I$, which means
  that, as $I$ increases, our ability to estimate the distribution
  improves, while accurately estimating the unnormalized intensity is
  more challenging. We chose to assume $I$ is known to discount this
  effect.}  The quality of a candidate estimator $f$ will be measured
in terms of the {\em risk}
$$
R(\tf,f) \deq \frac{1}{I^2} \left\| \tf - f \right \|^2_2.
$$

\subsection{Summary of notation}
\label{ssec:notation}

\rfm{Before proceeding to state and prove risk bounds for the proposed estimator, we summarize for the reader's convenience the principal notation used in the sequel:
\begin{itemize}
\item $m$: dimension of the original signal
\item $N (\ll m)$: number of measurements (detectors)
\item $\tf \in \reals^m_+$: unknown nonnegative-valued signal
\item $I = \sum_i \tf_i$: total intensity of $\tf$, assumed known
\item $Z \in \reals^{N \times m}$: random matrix with i.i.d.\ entries taking values $-\sqrt{(1-p)/p}$ with probability $p$ and $\sqrt{p/(1-p)}$ with probability $1-p$, where $p \in (0,1)$ is a tunable parameter
\item $\tA = Z/\sqrt{N}$: scaled matrix $Z$ (cf.~Theorem~\ref{thm:subgauss_bernoulli} for its norm preservation properties)
\item $\zeta_p$: subgaussianity constant of $\tA$, defined in \eqref{eq:zetap}
\item $c_1,c_2,c_3,c_4$: absolute constants from Theorem~\ref{thm:subgauss_bernoulli}
\item $A = \sqrt{p(1-p)/N} \tA + N^{-1} p(1-p) \1_{N \times m}$:  physically realizable sensing matrix, obtained by shifting and scaling of $\tA$; satisfies positivity and flux preservation requirements
\item $\Gamma \subset \reals^m_+$: finite or countably infinite set of candidate estimators with a penalty function $\pen : \Gamma \to \reals_+$ satisfying the Kraft inequality \eqref{eq:kraft}
\item $R(\tf,f) = \| \tf - f \|^2_2/I^2$: the risk of a candidate estimator $f$
\item $\hf$: the penalized maximum-likelihood estimator taking values in $\Gamma$, given by the solution to \eqref{eq:opt}
\end{itemize}
Other notation will be defined as needed in the appropriate sections.
}

\section{Bounds on the expected risk}
\label{sec:risk_bounds}

\rfm{Now we are in a position to establish risk bounds for the proposed estimator \eqref{eq:opt}. Theorem~\ref{thm:main} in Section~\ref{sec:oracle} is a general risk bound that holds (with high probability w.r.t.\ the realization of $\tA$) for any sufficiently regular class of candidate estimators and a suitable penalty functional satisfying the Kraft inequality. Section~\ref{sec:rates} then particularizes Theorem~\ref{thm:main} to the case in which the unknown signal $\tf$ is compressible in some known reference basis, and the penalty is proportional to the sparsity of a candidate estimator in the reference basis.}

\subsection{An oracle inequality for the expected risk}
\label{sec:oracle}

In this section we give an upper bound on
the expected risk $\expect R(\tf,\hf)$ that holds for any target
signal $\tf \succeq 0$ satisfying the normalization constraint
$\sum^m_{i=1} \tf_i = I$, without assuming anything about the sparsity
properties of $\tf$. Conceptually, our bound is an {\em oracle
  inequality}, which states that the expected risk of our estimator is
within a constant factor of the best regularized risk attainable by
estimators in $\Gamma$ with full knowledge of the underlying signal
$\tf$. More precisely, for each $f \in \Gamma$ define
$$
R^*(\tf,f) \deq  \frac{1}{I^2}\left\| \tf - f \right\|^2_2 + \frac{2\pen(f)}{I},
$$
and for every $\Upsilon \subseteq \Gamma$ define
$$
R^*(\tf,\Upsilon) \deq \min_{f \in \Upsilon} R^*(\tf,f),
$$
i.e.,~the best penalized risk that can be attained over $\Upsilon$ by an oracle that has full knowledge of $\tf$. We then have the following:

\begin{thm}\label{thm:main} Suppose that, in addition to the conditions stated in Section~\ref{sec:problem}, the feasible set $\Gamma$
  also satisfies the condition
\begin{equation}
A f \succeq (cI/N) \1_{N \times 1}, \qquad \forall f \in \Gamma 
\label{eq:positivity}
\end{equation}
for some $0 < c < 1$. Let $\cG_{N,p}$ be the collection of all subsets $\Upsilon \subseteq \Gamma$, such that $|\Upsilon| \le 2^{N/c_4 \zeta^4_p}$. Then the following holds with probability at least $1-me^{-KN}$ for some positive $K=K(c_1,c_3,p)$ (with respect to the realization of $\tA$):
\begin{equation}
\expect R(\tf,\hf) \le C_{N,p} \min_{\Upsilon \in \cG_{N,p}} R^*(\tf,\Upsilon) + \frac{2c^2_2 \zeta^4_p \log (c_2 \zeta^4_p m/N)}{N},
\label{eq:thm}
\end{equation}
where
$$
C_{N,p} \deq \max\left(\frac{24}{c}, \frac{16}{p(1-p)} \right)  N
$$
and the expectation is taken with respect to $y \sim \Poisson(A\tf)$.
\end{thm}

\begin{remark} {\em One way to satisfy the positivity condition
  (\ref{eq:positivity}) is to construct $\Gamma$ so that
\begin{align}\label{eq:positivity_0}
  f \succeq (cI) \1_{m \times 1}, \qquad \forall f \in \Gamma.
 \end{align}
  The desired condition \eqref{eq:positivity} then follows from \eqref{eq:pos_intens}. A condition similar to \eqref{eq:positivity_0} is natural in the context of estimating vectors with nonnegative entries from count data, as it
  excludes the possibility of assigning zero intensity to an input of
  a detector when at least one photon has been counted
  \cite{Csi91}.  }\hfil  \end{remark} 

\rmw{\begin{remark}{\em Both $C_{N,p}$ and
      $\zeta_p$ are minimized when $p=1/2$, suggesting that
      altering the sensing architecture to have $p \neq 1/2$ may impair
      performance, despite the fact that increasing $p$ would increase
      the expected total number of observed events (photons) and
      decreasing $p$ would decrease the DC offset of the measurements
      and hence measurement noise variance.}
\end{remark}}
\begin{remark} {\em Observe that for any pair $N_1 < N_2$ we have the
    inclusion $\cG_{N_1,p} \subseteq \cG_{N_2,p}$, which implies that
    $\min_{\Upsilon \in \cG_{N,p}}R^*(\tf,\Upsilon)$ is a {\em
      decreasing} function of $N$. Hence, the first term on the
    right-hand side of (\ref{eq:thm}) is the product of a quantity
    that increases with $N$ (i.e.,~$C_{N,p}$) and one that decreases
    with $N$. Combined with the presence of the $O(N^{-1}\log(m/N))$
    additive term, this points to the possibility of a {\em threshold
      effect}, i.e.,~the existence of a critical number of
    measurements $N^*$, below which the expected risk may not
    monotonically decrease with $N$ or $I$.}
\end{remark}

\subsection{Risk bounds for compressible signals}
\label{sec:rates}

We now use Theorem~\ref{thm:main} to analyze the
performance of the proposed estimator when the target signal $\tf$ is compressible (i.e.,~admits a sparse approximation) in some orthonormal reference basis.

Following \cite{CS:candes2}, we assume that there exists an orthonormal basis $\Phi = \{\phi_1,\ldots,\phi_m\}$ of $\reals^m$, such that $\tf$ is {\em compressible} in $\Phi$ in the following sense. Let $W$ be the orthogonal matrix with columns $\phi_1,\ldots,\phi_m$. Then the vector $\ttheta$ of the coefficients $\ttheta_j = \ave{\tf}{\phi_j}$ of $\tf$ in $\Phi$ is related to $\tf$ via $\tf = W\ttheta$. Let $\ttheta_{(1)},\ldots,\ttheta_{(m)}$ be the decreasing rearrangement of $\ttheta$: $|\ttheta_{(1)}| \ge |\ttheta_{(2)}| \ge \ldots \ge |\ttheta_{(m)}|$. We assume that there exist some $0 < q < \infty$ and $\rho > 0$, such that
\begin{equation}
|\ttheta_{(j)}| \le \rho I j^{-1/q}, \qquad j = 1,\ldots,m.
\label{eq:weak_lp}
\end{equation}
Note that for every $1 \le j \le m$ we have
$$
|\ttheta_{(j)}| \le \| \ttheta \|_2 = \| \tf \|_2 \le \| \tf \|_1 = I,
$$
so we can take $\rho$ to be a constant independent of $I$ or $m$. Any $\ttheta$ satisfying (\ref{eq:weak_lp}) is said to belong to the {\em weak-$\ell_q$ ball of radius $\rho I$}. The weak-$\ell_q$ condition (\ref{eq:weak_lp}) translates into the following approximation estimate \cite{CS:candes2}: given any $1 \le k \le m$, let $\thetak$ denote the best $k$-term approximation to $\ttheta$. Then
\begin{equation}
\frac{1}{I^2} \| \ttheta - \thetak \|^2_2 \le C \rho^2 k^{-2\alpha}, \qquad \alpha = 1/q - 1/2
\label{eq:compressibility}
\end{equation}
for some constant $C > 0$ that depends only on $q$. We also assume that $\tf$
satisfies the condition (\ref{eq:positivity_0}) for some $c \in (0,1)$, a
lower bound on which is assumed known.

\begin{thm}\label{thm:compressible} There exist a finite set of candidate estimators $\Gamma$ and a penalty function satisfying Kraft's inequality, such that the bound
\begin{align}\label{eq:compressible_risk}
& \expect R(\tf,\hf)  \le O(N) \min_{1 \le k \le k_*(N)} \left[ k^{-2\alpha} + \frac{k}{m} + \frac{k \log_2 m}{I}\right] \nonumber\\
& \qquad \qquad \qquad \qquad \qquad \qquad + O\left(\frac{\log(m/N)}{N}\right),
\end{align}
where
$$
k_*(N) \deq \frac{N}{2c_4 \zeta^4_p \log_2 m},
$$
holds with the same probability as in Theorem~\ref{thm:main}. The constants obscured by the $O(\cdot)$ notation depend only on $p$, $\rho$, $C$ and $c$.
\end{thm}

\noindent The proof is presented in Appendix~\ref{app:compressible_proof}; here we highlight a number of implications:

\noindent 1) In the low-intensity setting $I \le m\log m$, we get the risk bound
\begin{align*}
& \expect R(\tf,\hf) \le O(N) \min_{1 \le k \le k_*(N)} \left[k^{-2\alpha}  + \frac{2k \log_2 m}{I}\right]  \\
& \qquad \qquad \qquad \qquad \qquad+ O\left(\frac{\log(m/N)}{N}\right).
\end{align*}
If $k_*(N) \ge (\alpha I/\log_2 m)^{1/(2\alpha+1)}$, then we can further obtain
$$
\expect R(\tf,\hf) \le O(N) \left( \frac{I}{\log m} \right)^{-\frac{2\alpha}{2\alpha + 1}} + O\left(\frac{\log(m/N)}{N}\right).
$$
If $k_*(N) < (\alpha I/\log_2 m)^{1/(2\alpha +1)}$, there are not enough measurements, and the estimator saturates, although its risk can be controlled.

\noindent 2) In the high-intensity setting $I > m \log m$, we obtain\begin{align*}
& \expect R(\tf,\hf) \le
O(N) \min_{1 \le k \le k_*(N)} \left[k^{-2\alpha} + \frac{2k}{m}\right] \\
& \qquad \qquad \qquad \qquad \qquad + O\left(\frac{\log(m/N)}{N}\right).
\end{align*}
If $k_*(N) \ge (\alpha m)^{1/(2\alpha+1)}$, then we can further get
$$
\expect R(\tf,\hf) \le O(N) m^{-\frac{2\alpha}{2\alpha + 1}} +O\left(\frac{\log(m/N)}{N}\right).
$$
Again, if $k_*(N) < (\alpha m)^{1/(2\alpha+1)}$, there are not enough measurements, and the estimator saturates. 

\noindent 3) When $I \asymp m$ and $N \asymp m^{1/\beta}$ for some $\beta > 1 + 1/2\alpha$, we get (up to log terms) the rates
$$
\expect R(\tf,\hf) = O \left(m^{-\gamma}\right),
$$
where $\gamma = \frac{2\alpha - (2\alpha + 1)/\beta}{2\alpha + 1} > 0$.

\section{Experimental Results}
\label{sec:experiments}

In this section we present the results of a proof-of-concept
experiment showing the effectiveness of sparsity-regularized
Poisson log likelihood reconstruction from CS measurements. 
In previous work \cite{harmanyPCS}, we described an optimization
formulation called {\em SPIRAL (Sparse Poisson Intensity Reconstruction Algorithms)} for solving a simpler variant of (\ref{eq:CSP2}): 
\begin{equation}
\widehat{f} = \argmin_f \left[ \phi(f) + \tau \pen(f) \right] \ \text{subject to} \ f \succeq 0,
\label{eq:SPIRAL}
\end{equation}
where $\phi(f) =  \sum_j (Af)_j - y_j \log (Af)_j$.
In our setup, since $A$ has nonnegative entries, the constraint $Af \succeq 0 $
in (\ref{eq:CSP2}) is redundant.  Additionally, we do not require that the total intensity of
the reconstruction $f$ must sum to $I$ since the resulting problem is easier to solve,
and this equality constraint, in general, is approximately satisfied
at the solution, i.e., it is not necessary to obtain accurate experimental results.

The proposed approach sequentially approximates 
the objective function in (\ref{eq:SPIRAL}) by separable quadratic
problems (QP) that are easier to minimize.  
In particular, at the $k$-th iteration we use
the second-order Taylor expansion of $\phi$ around $f^k$ and
approximate the Hessian by a positive scalar ($\eta_k$) multiple of the 
identity matrix, resulting in the following minimization problem:
\begin{eqnarray}
\displaystyle \nonumber
	f^{k+1}  =  \displaystyle \argmin_{f} 
	&& \!\! \displaystyle \left  \| \left [ \! f^k \!-\! \frac{1}{\eta_k} \nabla \phi(f^k) \! \right ]- f \right  \|_2^2
	\!\!\!+\! \displaystyle \frac{2\tau}{\eta_k}  \pen( f ) \\
	\textrm{subject to} \!\!\! \! \! && \!\! f \succeq 0, \;        \label{eq:sqp}
\end{eqnarray}
which can be viewed as a denoising subproblem applied to the gradient descent. This gives us considerable flexibility in designing effective penalty functions
and in particular allows us to incorporate sophisticated ``sparsity
models'' which encode, for instance, persistence of significant
wavelet coefficients across scales to improve reconstruction
performance. \rfm{In the experiments below we utilize one such penalty,
a partition-based estimator, as described in \cite{willett:riskb}.}

\rfm{Partition-based methods calculate image
estimates by determining the ideal partition of the domain
of observations and by using maximum likelihood estimation
to fit a model (e.g., a constant) to each cell in the optimal
partition. The space of possible partitions is a nested hierarchy
defined through a recursive dyadic partition (RDP) of the
image domain, and the optimal partition is selected by pruning
a quad-tree representation of the observed data to best fit the
observations with minimal complexity. We call this
partition-based algorithm SPIRAL-RDP.
An additional averaging-over-shifts 
(cycle spinning) step  can be efficiently incorporated to yield a 
\emph{translationally-invariant} (TI) algorithm, which we call SPIRAL-RDP-TI, 
that results in more accurate reconstructions.  }

\rfm{The main computational costs of the SPIRAL methods come from
  matrix-vector multiplications involving $A$ for calculating $\eta_k$
  and $\nabla \phi(x_k)$ in (\ref{eq:sqp}).  Specifically, at each
  iteration $k$, SPIRAL computes two matrix-vector multiplications
  each with $A$ and with $A^T$. For SPIRAL-RDP and SPIRAL-RDP-TI, even
  though the partition-based penalty QP appears difficult to solve
  because of its nonconvexity due to the penalty term, its global
  minimizer is easily computed using a non-iterative tree-pruning
  algorithm (see \cite{harmanyPCS} and \cite{willett:riskb} for
  details).  }

\begin{figure*}[t!]
\centering
\begin{tabular}{ccc}  
	\includegraphics[width=5.6cm]{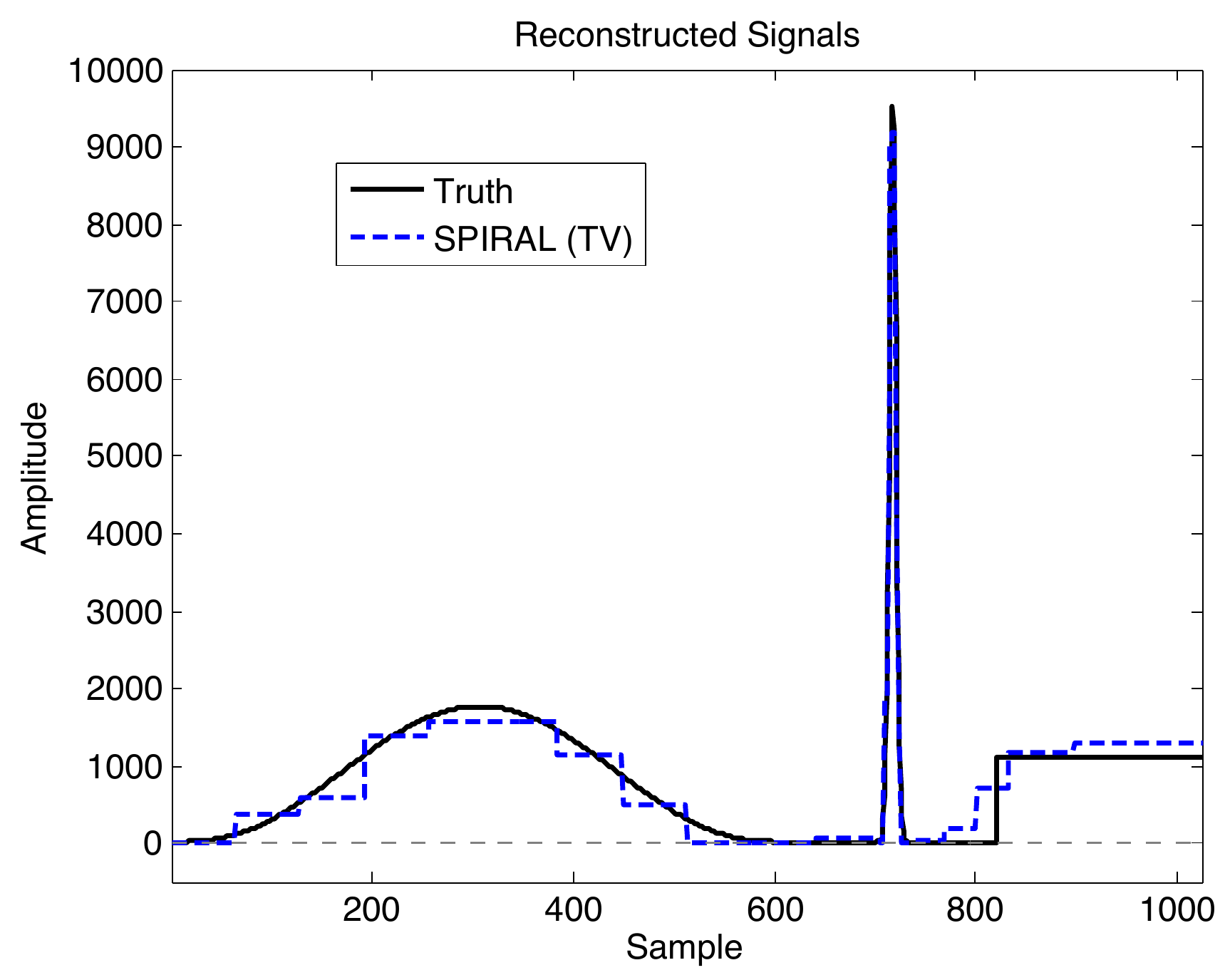} &
	\includegraphics[width=5.6cm]{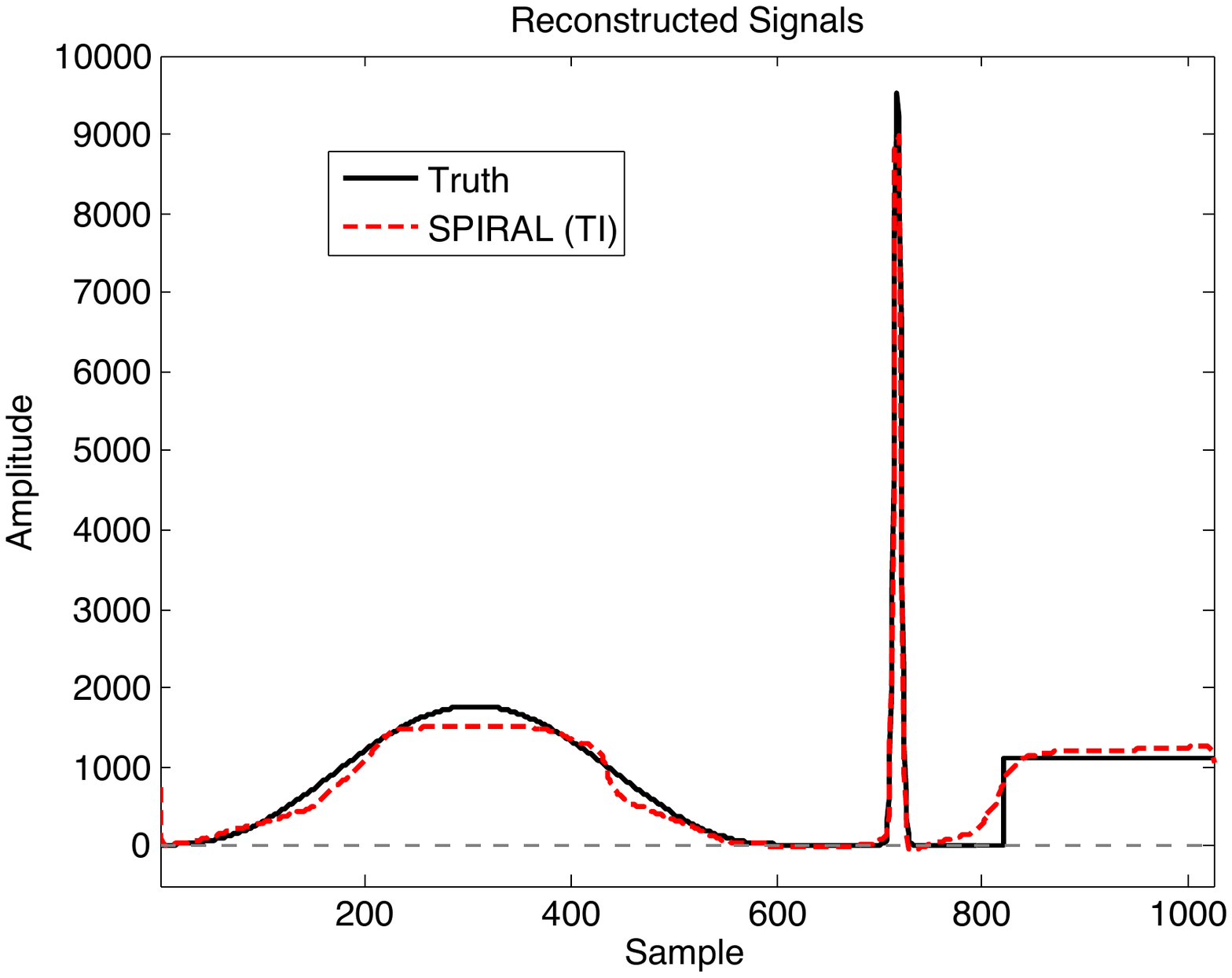} &
	\includegraphics[width=5.15cm]{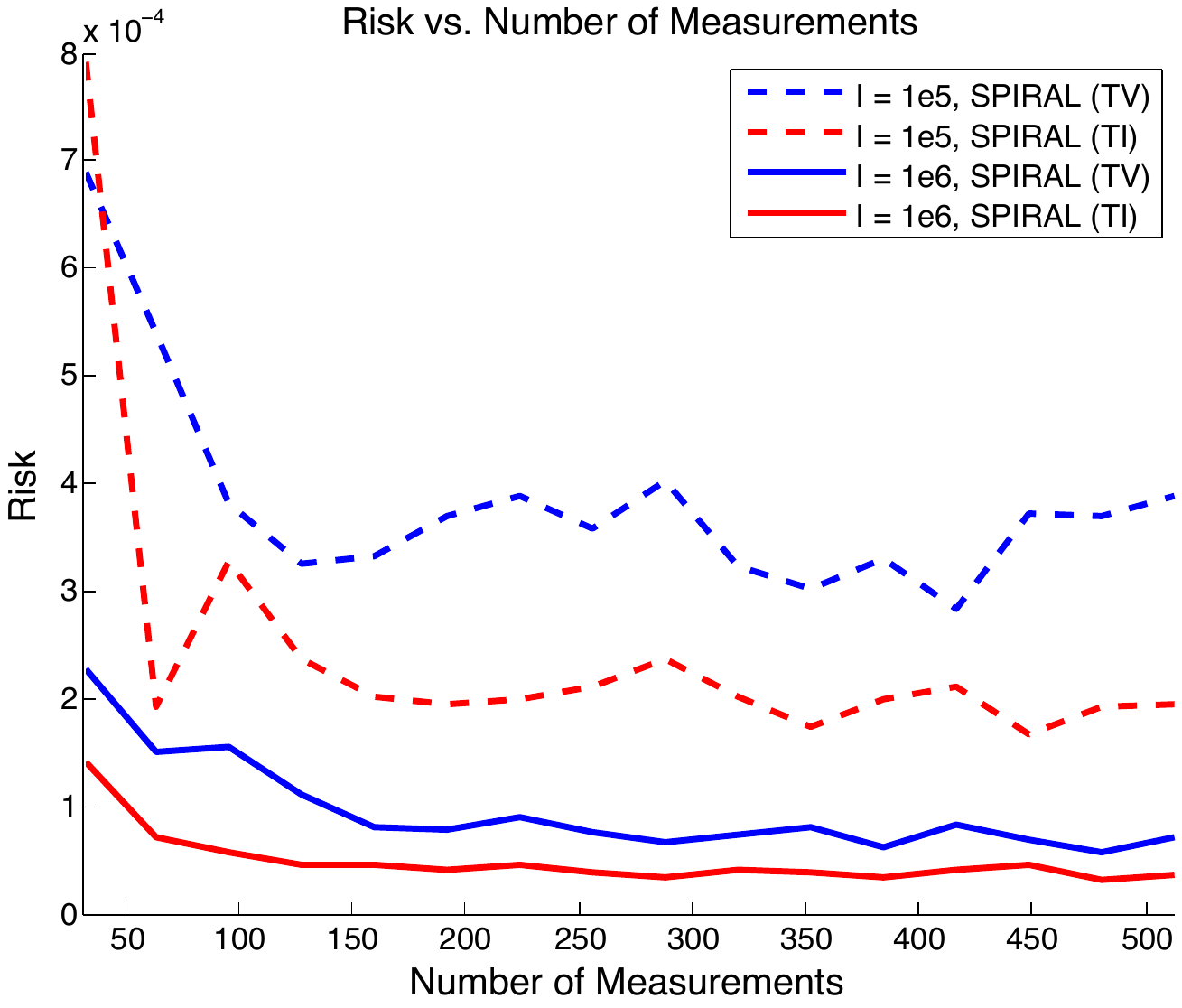}\\
	\qquad (a) & \qquad (b) & \qquad (c)  
  \end{tabular}
  \caption{\rfm{Reconstruction results using the SPIRAL algorithm.
      (a) SPIRAL-RDP
      yields a nonsmooth piecewise constant estimate $\hf_{\rm RDP}$
      with a corresponding risk $R(\hf_{\rm RDP}, \tf) =
      7.552\times10^{-5}$.  (b) SPIRAL-RDP-TI produces a
      smoother and more accurate reconstruction $\hf_{\rm RDP-TI}$ with
      corresponding risk $R(\hf_{\rm RDP-TI}, \tf) = 4.468\times 10^{-5}$.
      The plot (c) shows the reconstruction accuracy versus number of
      measurements for two particular values of the intensity $I$
      ($10^5$ and $10^6$).}}
  \label{fig:results}
 \end{figure*}

We evaluate the effectiveness of the proposed approaches in
reconstructing a piecewise smooth function from noisy compressive
measurements.  In our simulations, the true signal (the black line in
\rfm{Figs.~\ref{fig:results}(a) and \ref{fig:results}(b)}) is of length 1024
\rfm{and total intensity $I = 8.2 \times 10^{5}$}.  We take 512
noisy compressive measurements of the signal using a sensing matrix
that contains 32 randomly distributed nonzero elements per row.  This
setup yields a mean detector photon count of 50, ranging from as few
as 22 photons, to as high as 94 photons.  We allowed each algorithm a
fixed time budget of three seconds in which to run, which is
sufficient to yield approximate convergence for all methods
considered.  Each algorithm was initialized at the same starting
point: if $z = A^Ty$, and $x : x_i = y_i/(Az)_i$, then we initialize
with $f^0 : f_i^0 = z_i (A^Tx)_i/(A^T \1)_i$, where $\1$ is
a vector of ones.  The value of the
regularization parameter $\tau$ was tuned independently for each
algorithm to yield the minimal risk 
$R(\hf,\tf) = \|\hf - \tf\|_2^2/I^2$
at the exhaustion of the computation budget.
\rfm{Tuning the regularization parameter in this manner
is convenient in a simulation study.  However, in the absence of truth, 
one typically resorts to a cross-validation procedure to determine an appropriate
level of regularization.}

\sloppypar{\rfm{In Fig.~\ref{fig:results}(a), we see that models
    within a partition (constant pieces) are less smooth than that of
    the original signal; however this drawback can be effectively
    neutralized through cycle spinning (see
    Fig.~\ref{fig:results}(b)).  In addition, the accuracy of the
    reconstruction (measured using the risk $R(\hf, \tf)$) is improved
    by this averaging of shifts. Specifically, the SPIRAL-RDP estimate
    $\hf_{\rm RDP}$ has a risk of $R(\hf_{\rm RDP}, \tf) = 7.552\times
    10^{-5}$, while the SPIRAL-RDP-TI estimate $\hf_{\rm RDP-TI}$
    achieves a much lower risk of $R(\hf_{\rm RDP-TI},\tf) = 4.468
    \times 10^{-5}$. In Fig.~\ref{fig:results}(c), we examine how the
    performance of both partition-based SPIRAL methods scale with the
    number of measurements.  These results utilize the same true
    signal and sensing matrix type as before, and are averaged over
    four noise realizations.  By choosing two different intensity
    levels, we see that a higher intensity consistently leads to
    better performance.  However, for a fixed intensity, the benefits
    of a higher number of measurements are less pronounced since
    collecting more observations necessarily results in a lower
    intensity per observation and hence noisier measurements (i.e.,
    less photons collected per measurement).  Note that the plot in
    Fig.~\ref{fig:results}(c) is not smoothly decreasing as one would
    expect; as we change the number of measurements, we need to
    randomly generate new Poisson realizations of our data, and hence
    there is some degree of variability in these results.}

\rfm{In \cite{harmanyPCS}, we examine our SPIRAL approach 
with an $\ell_1$-norm  penalty on the coefficients of a wavelet expansion of the signal. In
this case, the resulting reconstruction is very oscillatory with pronounced wavelet
artifacts.  With an increase in the regularization parameter these artifacts can be
minimized; however, this leads to an ``oversmoothed'' solution and increases the
risk of the estimate. In addition, we compare the SPIRAL approaches to the 
more established  expectation-maximization algorithms based upon the same 
maximum penalized likelihood estimation objective function 
found in \eqref{eq:sqp} and demonstrate that reconstructions from the partition-based
SPIRAL methods are more accurate. In particular, they produce
estimates with lower risk values, are more effective in recovering regions of low intensity, 
and yield reconstructions without spurious wavelet artifacts.}

\rfm{We mention other recent approaches for solving Poisson inverse
  problems; given that a detailed comparison of the performances of
  these various methods is beyond the scope of this paper, we simply
  outline some potential drawbacks with these approaches.  In
  \cite{starckPoissonCS}, the Poisson statistical model is bypassed in
  favor of an additive Gaussian noise model through the use of the
  Anscombe variance-stabilizing transform. This statistical
  simplification is not without cost, as the linear projections of the
  scene must now be characterized as nonlinear observations.  Other
  recent efforts \cite{marioPCS,setzer} solve Poisson image
  reconstruction problems with total variation norm regularization,
  but the method involves a matrix inverse operation, which can be
  extremely difficult to compute for large problems outside of
  deconvolution settings. Finally, the approaches in
  \cite{combettes,chaux} apply proximal functions to solve more
  general constrained convex minimization problems.  These methods use
  projection to obtain feasible iterates (i.e., nonnegative intensity
  values), which may be difficult for recovering signals that are
  sparse in a noncanonical basis.}

\section{Conclusions}
\label{sec:conclude}

We have derived upper bounds on the compressed sensing estimation
error under Poisson noise for sparse or compressible signals. We
specifically prove error decay rates for the case where the penalty
term is proportional to the $\ell_0$-norm of the solution; this form
of penalty has been used effectively in practice with a
computationally efficient Expectation-Maximization algorithm
(cf.~\cite{gehm:ddis}), but was lacking the theoretical support
provided by this paper. Furthermore, the main theoretical result of
this paper holds for any penalization scheme satisfying the Kraft
inequality, and hence can be used to assess the performance of a
variety of potential reconstruction strategies besides
sparsity-promoting reconstructions.

One significant aspect of the bounds derived in this paper is that their signal-dependent portion grows with $N$, the size of the measurement array, which is a
major departure from similar bounds in the Gaussian or bounded-noise
settings. It does not appear that this is a simple artifact of our
analysis. Rather, this behavior can be intuitively understood to
reflect that elements of $y$ will all have similar values at low light levels,
making it very difficult to infer the relatively small variations in
$\tA\tf$.  Hence, compressed sensing using shifted Rademacher sensing
matrices is fundamentally difficult when the data are Poisson
observations.
\begin{appendices}

\section{Proof of Theorem~\ref{thm:subgauss_bernoulli}}
\label{app:subgaussian_proof}

The proof makes heavy use of the geometric approach of
\cite{MenPajTom07,MenPajTom08}. Since this approach is not as
well-known in the compressed sensing community as the usual RIP, we
give a brief exposition of its main
tenets. Consider the problem of recovering an unknown signal $\tf$,
which resides in some set $\Lambda \subset \reals^m$, from $N$ linear
measurements of the form $y_1 = \dotprod{Z_1}{\tf}, \ldots, y_N =
\dotprod{Z_N}{\tf}$, where the measurement vectors $Z_1,\ldots,Z_N \in
\reals^m$ are i.i.d.\ samples from a distribution $\mu$ which is:
\begin{itemize}
\item {\em subgaussian with constant $\zeta$}, i.e., there exists a constant $\zeta > 0$, such that for $Z_0 \sim \mu$ and for every $u \in \reals^m$,
\begin{equation}\label{eq:orlicz}
\inf \left\{ s \ge 0 : \expect \exp \left( \frac{|\dotprod{Z_0}{u}|^2}{s^2} \right) \le 2 \right\} \le \zeta \| u \|_2
\end{equation}
\item {\em isotropic}, i.e., for $Z_0 \sim \mu$ and for every $u \in \reals^m$, $\expect |\dotprod{Z_0}{u}|^2 = \| u \|^2_2$.
\end{itemize}
The main results  of \cite{MenPajTom07} revolve around the norm preservation properties of the {\em random operator} $\tA : \reals^m \to \reals^N$ defined by
\begin{equation}\label{eq:random_op}
\tA u \deq \frac{1}{\sqrt{N}} \sum^N_{i=1} \dotprod{Z_i}{ u} e_i,
\end{equation}
where $e_1,\ldots,e_N$ is the standard basis in $\reals^N$. Particularized to the case of $\Lambda = B^m_1$, the unit ball w.r.t.\ the $\ell_1$ norm on $\reals^m$, the first main result of \cite{MenPajTom07} reads as follows:
\begin{thm}\cite[Theorem~A]{MenPajTom07}\label{thm:subgauss_1} Let
  $Z_1,\ldots,Z_N \in \reals^m$ be i.i.d.\ samples from an ensemble
  $\mu$ which is isotropic and subgaussian with constant $\zeta \ge
  1$. There exist absolute constants $c_1,c_2 > 0$, such that, with
  probability at least $1-e^{-c_1 N/\zeta^4}$,
\begin{equation}
\| u - v \|_2^2 \le 4 \|\tA (u-v) \|_2^2 +  \frac{2 c_2^2 \zeta^4 \log (c_2 \zeta^4 m/N)}{N}
\label{eq:subgauss_1b}
\end{equation}
for all $u,v \in B^m_1$.
\end{thm}
The other main result of \cite{MenPajTom07}, informally, states the
following: for any finite set $T \subset S^{m-1}$, the
random operator $\tA$ does not distort the norms of the elements of
$T$ too much, provided the number of measurements $N$ is sufficiently
large. The required minimal number of measurements is determined by
the cardinality of $T$. In its precise form, the relevant result of
\cite{MenPajTom07} says the following:
\begin{thm}\cite[Corollary 2.7]{MenPajTom07}\label{thm:subgauss_2} 
 There exist absolute constants $c_3,c_4 > 0$, such that the
  following holds. Consider a finite set $T \subset S^{m-1}$, and let
  $Z_1,\ldots,Z_N \in \reals^m$ be i.i.d.\ samples from a
  $\zeta$-subgaussian isotropic ensemble.  Then, with probability at
  least $1-e^{-c_3 N/\zeta^4}$,
\begin{equation}
  \frac{1 }{2} \le \| \tA u\|_2^2 \equiv \frac{1}{N}\sum^N_{i=1}
  |\dotprod{Z_i}{u}|^2 \le \frac{3}{2}, \qquad \forall u \in \Lambda 
\label{eq:subgauss_2b}
\end{equation}
provided $N \ge c_4 \zeta^4 \log_2 |\Lambda|$.
\end{thm}
\begin{remark} {\em Theorem~\ref{thm:subgauss_2} is a special case of a more general result that applies to {\em general} (not necessarily finite) subsets $T$ of the unit sphere. The minimum necessary number of measurements is determined by the so-called {\em $\ell_*$-functional} of $T$, which is defined as follows. Let $g_1,\ldots,g_m$ be independent standard Gaussian random variables, i.e., each $g_i \sim N(0,1)$ independently of all others. Then
$$
\ell_*(T) \deq \expect \sup_{u \in T} \left| \sum^m_{i=1} g_i u_i \right|,
$$
where the expectation is taken w.r.t.\ $g_1,\ldots,g_n$, and $u_i$
denotes the $i$th component of $u$. Informally, $\ell_*(T)$ measures
how much the elements of $T$ are correlated with white Gaussian
noise. Estimates of $\ell_*(T)$ are available for many sets $T$. For
instance (see Section~3 of \cite{MenPajTom07} and references therein):
\begin{itemize}
\item If $T$ is a finite set, then $\ell_*(T) \le c \sqrt{\log_2 |T|}$
  for some absolute constant $c > 0$. 
\item If $T$ is the set of all convex combinations of unit vectors in
  $\reals^m$ whose $\ell_0$ norm is at most $k$, 
\begin{equation}
T =  \text{conv\,hull}\left\{ u \in S^{m-1} : \| u \|_0 = |\{ i : u_i \neq 0 \}| \le k
\right\}, 
\label{eq:short_support}
\end{equation}
then $\ell_*(T) \le c \sqrt{k \log_2 (cm/k)}$ for some absolute
constant $c > 0$. (We do not use this particular $T$ in our
analysis, but mention it here because of its connection to the more
widely known RIP \cite{RIP}.)
\end{itemize}
The minimum necessary number of measurements for (\ref{eq:subgauss_2})
to hold with the same probability as before is determined by $N \ge c_4 \zeta^4
\ell_*(T)^2$. When $|T|$ is finite, combining this bound with the
estimate of $\ell_*(T)$ in terms of the log-cardinality of $T$ yields
Theorem~\ref{thm:subgauss_2}. Moreover, as shown in
\cite{MenPajTom08}, the usual RIP for matrices with rows drawn from
subgaussian isotropic ensembles is a consequence of this result as
well. Specifically, it relies on the $\ell_*(T)$ estimate for the set
$T$ defined in (\ref{eq:short_support}).}
\end{remark}

We now apply Theorems~\ref{thm:subgauss_1} and \ref{thm:subgauss_2} to
the measurement matrix $\tA$ defined in Section~\ref{sec:A_prop}. Recall that $\tA = Z/\sqrt{N}$, and let $Z_i = (Z_{i,1},\ldots,Z_{i,m})$
denote the $i$th row of $Z$. By construction, each $Z_i$ is an
independent copy of a random variable $Z_0 \in \reals^m$ with
distribution $\nu^{\otimes m}_p$ --- that is, the components of $Z_0$
are drawn i.i.d.\ from $\nu_p$. To be able to apply
Theorems~\ref{thm:subgauss_1} and \ref{thm:subgauss_2}, we first show
that the distribution $\nu^{\otimes m}_p$ is subgaussian and
isotropic. To that end, we need to obtain a bound of the form
(\ref{eq:orlicz}) for linear functionals of the form
$\dotprod{Z_0}{u}$. The infimum on the left-hand side of
Eq.~(\ref{eq:orlicz}) is the so-called {\em Orlicz $\psi_2$-norm} of
the random variable $\dotprod{Z_0}{u}$. Given a real-valued random variable $U$, its Orlicz $\psi_2$-norm \cite[Ch.~2]{WaaWel96} is defined as
$$
\| U \|_{\psi_2} \deq \inf \left\{ s \ge 0 : \expect \exp \left( \frac{U^2}{s^2} \right) \le 2 \right\}.
$$
Thus, $\mu$ is subgaussian with constant $\zeta$ if for $Z_0 \sim \mu$ we have
$$
\| \dotprod{Z_0}{u} \|_{\psi_2} \le \zeta \| u \|_2, \qquad \forall u \in \reals^m.
$$
Here is a useful tool for bounding Orlicz norms:
\begin{lemma}\cite[Lemma~2.2.1]{WaaWel96}\label{lm:orlicz_bound} Let $U$ be a real-valued random variable that satisfies the tail bound
$$
\prob \left[ |U| > t \right] \le Ke^{-Ct^2}
$$
for all $t > 0$, where $K,C > 0$ are some constants. Then its Orlicz $\psi_2$-norm satisfies $\| U \|_{\psi_2} \le \sqrt{(1+K)/C}$.
\end{lemma}
Using this lemma, we can prove the following:

\begin{lemma}\label{lm:skew_subgauss} The product probability measure
  $\nu^{\otimes m}_p$ is isotropic and subgaussian with constant 
  $\zeta_p$ defined in (\ref{eq:zetap}).\end{lemma}

\begin{proof} Let $Z_0 = (Z_{0,1},\ldots,Z_{0,m}) \sim \nu^{\otimes
    m}_p$. Isotropy is immediate from (\ref{eq:isotropy}). To prove subgaussianity, let us first assume that $p\neq 1/2$. Fix
some $u \in \reals^m$ and consider the random variable
$\dotprod{Z_0}{u}$, which is a sum of independent random variables
$Z_{0,j}u_j$, $1 \le j \le m$. Then $\expect \dotprod{Z_0}{ u} =
0$, and with probability one each $Z_{0,j} u_j$ takes values in
the set $\{-\lambda^-_p |u_j|,-\lambda^+_p |u_j|\}$ if $u_j < 0$, and
in $\{\lambda^-_p |u_j|, \lambda^+_p |u_j| \}$ if $u_j  \ge
0$. Hence, Hoeffding's inequality gives the tail bound
\begin{align*}
\prob \big[ |\dotprod{Z_0}{ u} | > t \big] &\le 2 \exp\left(-\frac{2t^2}{(\lambda^+_p - \lambda^-_p)^2 \sum^m_{j=1} |u_j|^2}\right) \\
& = 2 \exp \left(-\frac{2 p(1-p) t^2}{\| u \|^2_2} \right).
\end{align*}
Using Lemma~\ref{lm:orlicz_bound} with $K = 2$ and $C =
2p(1-p)/\|u\|^2_2$, we get the desired result. When $p=1/2$, using the
fact that the Rademacher measure is symmetric, it can be shown that
$\nu^{\otimes m}_{1/2}$ is subgaussian with constant $\zeta = 1$
\cite{MenPajTom07}.
\end{proof}

Now, using
  Theorems~\ref{thm:subgauss_1} and \ref{thm:subgauss_2} in
  conjunction with Lemma~\ref{lm:skew_subgauss}, we have proved
  Theorem~\ref{thm:subgauss_bernoulli}.
  
  \section{Proof of Theorem~\ref{thm:main}}
  \label{app:oracle_proof}
  
With probability at least $1-e^{-c_3 N/\zeta^4_p}$, the following chain of estimates holds:
\begin{eqnarray*}
\lefteqn{\frac{1}{I^2} \| \tf - \hf \|^2_2} \\
&\le& \frac{4}{I^2} \| \tA (\tf - \hf) \|^2_2 + \frac{2c^2_2 \zeta^4_p \log (c_2 \zeta^4_p m/N)}{N}  \\
&=& \frac{4N}{p(1-p)I^2} \| A (\tf - \hf) \|^2_2 + \frac{2c^2_2 \zeta^4_p \log (c_2 \zeta^4_p m/N)}{N}   \\
&\le& \frac{4N}{p(1-p)I^2} \| A (\tf - \hf) \|^2_1 + \frac{2c^2_2 \zeta^4_p \log (c_2 \zeta^4_p m/N)}{N},
\end{eqnarray*}
where the first inequality is a consequence of the first part of
Theorem~\ref{thm:subgauss_bernoulli}, and the remaining steps follow
from definitions, standard inequalities for $\ell_p$ norms, and the
fact that $\sum_i f_i = \sum_i \tf_i = I$ for all $f \in
\Gamma$ \footnote{\rmw{We use the fact that
$\|A(f^* - \hat{f})\|_2^2 = \frac{N}{p(1-p)}\|\widetilde{A}(f^* -
\hat{f})\|_2^2$, which is only true when $I = \|f^*\|_1 =
\|\hat{f}\|_1$. If we did not assume $I$ was known, we would have
additional terms in our error bound which would be proportional to $I$
and would not reflect our ability to exploit prior knowledge of
structure or sparsity to achieve an accurate solution.}}. Moreover,
\begin{eqnarray*}
\lefteqn{\| A (\tf - \hf) \|^2_1 } \nonumber \\
&=& \left(\sum^N_{i=1} \left| (A\tf)^{1/2}_i - (A\hf)^{1/2}_i \right|
  \cdot \left|(A\tf)^{1/2}_i +  (A\hf)^{1/2}_i \right|   \right)^2
\nonumber \\
&\le& \sum^N_{i,j=1} \left|(A\tf)^{1/2}_i - (A\hf)^{1/2}_i \right|^2
\left|(A\tf)^{1/2}_j +  (A\hf)^{1/2}_j \right|^2 \nonumber \\
&\le& 2 \sum^N_{i,j=1} \left|(A\tf)^{1/2}_i - (A\hf)^{1/2}_i \right|^2
\cdot \left( (A\tf)_j + (A\hf)_j \right) \nonumber \\
&\le& 4 I \sum^N_{i=1} \left|(A\tf)^{1/2}_i - (A\hf)^{1/2}_i
  \right|^2, 
\end{eqnarray*}
where the first inequality follows from Cauchy--Schwarz, the second is
a consequence of the arithmetic-mean/geometric-mean inequality, and the third follows from (\ref{eq:flux}). It is a matter of
straightforward algebra (see Appendix~\ref{app:affinity} below) to show that
\begin{eqnarray}
\lefteqn{\sum^N_{i=1} \left|(A\tf)^{1/2}_i - (A\hf)^{1/2}_i \right|^2} \nonumber \\
&=& -2 \log \prod^N_{i=1} \exp \left(-\frac{1}{2} \left[ (A\tf)^{1/2}_i - (A\hf)^{1/2}_i \right]\right)^2 \nonumber \\
&=& 2 \log \frac{1}{\int \sqrt{ p(y|A\tf) p(y|A\hf)} d\nu(y)}, \label{eq:affinity}
\end{eqnarray}
where $\nu$ is the counting measure on $\integers_+^N$. Now,  the same techniques as in Li and Barron \cite{libarron} (see also the proof of Theorem~7 in \cite{kola2} or Appendix~\ref{app:li_barron} below) can be used to show that 
\begin{eqnarray}
\lefteqn{2 \expect \log \frac{1}{\int \sqrt{ p(y|A\tf) p(y|A\hf)} d\nu(y)} }\nonumber \\
&\le& \min_{f \in \Gamma} \left[ \KL\Big(p(\cdot|A\tf) \Big\| p(\cdot|Af)\Big) + 2\pen(f) \right],
\label{eq:li_barron}
\end{eqnarray}
where $\KL(\cdot \| \cdot)$ is the {\em Kullback--Leibler (KL)
  divergence}, which for the Poisson likelihood has the form
\begin{eqnarray*}
\lefteqn{\KL\Big(p(\cdot|A\tf) \Big\| p(\cdot|Af)\Big)} \\
& =& \sum^N_{i=1} \left[(A\tf)_i \log \frac{(A\tf)_i}{(Af)_i} - (A\tf)_i + (Af)_i\right].
\end{eqnarray*}
Using the inequality $\log t \le t -1$ together with (\ref{eq:positivity}), which holds with probability at least $1-Np^m$, and (\ref{eq:pos_intens}), we can bound the KL divergence as
\begin{eqnarray*}
\lefteqn{\sum^N_{i=1} \left[(A\tf)_i \log \frac{(A\tf)_i}{(Af)_i} - (A\tf)_i + (Af)_i\right]} \\
&\le& \sum^N_{i=1} \left[(A\tf)_i \left( \frac{(A\tf)_i}{(Af)_i} - 1\right) - (A\tf)_i + (Af)_i\right] \\
&=& \sum^N_{i=1} \frac{1}{(Af)_i} \left[ (Af)^2_i - 2(Af)_i (A\tf)_i + (A\tf)^2_i\right] \\
&\le& \frac{N}{cI} \| A(\tf - f) \|^2_2 \\
&=& \frac{p(1-p)}{cI} \| \tA (\tf - f ) \|^2_2.
\end{eqnarray*}
Now, choose any $\Upsilon^* \in \cG_{N,p}$, such that
$$
R^*(\tf,\Upsilon^*) = \min_{\Upsilon \in \cG_{N,p}} R^*(\tf,\Upsilon).
$$
Observe that $N \ge c_4 \zeta^4_p \log|\Upsilon^*|$, so we can apply the second part of Theorem~\ref{thm:subgauss_bernoulli} to the set $\left\{ \frac{\tf - f}{\| \tf - f \|_2} : f \in \Upsilon^*\right\}$. Then, with probability at least $1-e^{-c_3 N/\zeta^4_p}$, we have
$$
\| \tA(\tf - f) \|^2_2 \le (3/2) \| \tf - f \|^2_2, \qquad \forall f \in \Upsilon^*.
$$
Combining everything and rearranging, we get the bound
\begin{align*}
& \expect R(\tf,\hf) \le \\
& C_{N,p} \min_{f \in \Upsilon^*} \left[\left\| \frac{\tf}{I} - \frac{f}{I} \right\|^2_2 + \frac{2\pen(f)}{I} \right] + \frac{2c^2_2 \log (c_2 \zeta^4_p m/N)}{N}
\end{align*}
which holds with probability at least $1-(Np^m + e^{-c_1 N/\zeta^4_p} + e^{-c_3 N/\zeta^4_p}) \ge 1 - me^{-KN}$ for a suitable constant $K = K(c_1,c_3,p)$. The theorem is proved.

\section{Proof of Theorem~\ref{thm:compressible}}
\label{app:compressible_proof}

In order to apply Theorem~\ref{thm:main}, we will form a
suitable finite class of estimators $\Gamma$ and set a penalty function
$\pen(f)$ over this class which (a) is smaller for sparser
$\theta = W^T f$ and (b) satisfies (\ref{eq:kraft}).  The family $\Gamma$ is
constructed as follows.
\begin{enumerate}
\item Define the sets
\begin{eqnarray*}
  \lefteqn{ \Theta \deq \left\{ \theta \in \reals^m : \| \theta \|_\infty
      \le I; \right. } \\
  &&   \mbox{ each $\theta_i$ uniformly quantized into one of}
    \\
&& \left. \mbox{ $\sqrt{m}$
      bins over $[-I,I]$} \right\} 
\end{eqnarray*}
and $\cF \deq \left\{ f = W\theta : \theta \in \Theta \right\}$.
\item For each $f \in \cF$, let $\bar{f}$ denote its $\ell_2$
  projection  onto the closed convex set
  $$
  \cC \deq \left\{ g \in \reals^m: g \succeq (cI)\1_{m \times 1} \mbox{ and } \sum^m_{i=1} g_i = I \right\},
  $$
i.e.,
$$
\bar{f} \deq \argmin_{g \in \cC} \| f - g \|_2.
$$
\item Finally, let $\Gamma \deq \left\{ \bar{\theta} = W^T \bar{f} : f \in \cF \right\}$.
\end{enumerate}
Note that the projection $\bar{f}$ satisfies the {\em Pythagorean identity}
$$
\| g - f \|^2_2 \ge \| g - \bar{f} \|^2_2 + \| \bar{f} - f \|^2_2, \qquad \forall g \in \cC
$$
(see, e.g.,~Theorem~2.4.1 in \cite{CenZen97}). In particular, $\| g - f \|^2_2 \ge \| g - \bar{f} \|^2_2$, and, since $\tf \in \cC$, we have
\begin{equation}
\| \tf - \bar{f} \|^2_2 \le \| \tf - f \|^2_2, \qquad\forall f \in \cF.
\label{eq:projection}
\end{equation}

Consider the penalty
$$
\pen(f) = \log_2(m+1) +  (3/2) \|\theta\|_{0} \log_2(m) , \qquad \theta = W^T f
$$
(measured in bits; in order to satisfy Kraft's inequality as stated in (\ref{eq:kraft}), it will need to be rescaled by $\log 2$). This corresponds to the following prefix code for $\theta \in \Theta$
(that is, we encode the elements of $\Theta$, before they are
subjected to the deterministic operation of projecting onto $\cC$):
\begin{enumerate}
\item First we encode $\| \theta \|_0$, the number of nonzero components of $\theta$, which can
  be encoded with $\log_2(m+1)$ bits.
\item For each of the $\|\theta\|_{0}$ nonzero components, we encode
  its location in the $\theta$ vector; since there are $m$ possible
  locations, this takes $\log_2(m)$ bits per component.
\item Next we encode each coefficient value, quantized to one of
  $\sqrt{m}$ uniformly sized bins.
\end{enumerate}
Since this corresponds to a uniquely decodable code for $f \in \cF$ (or $\theta \in \Theta)$, we see that $\pen(f)$ indeed satisfies
the Kraft inequality \cite{CovTho06}.

Now, given $\ttheta = W^T \tf$, let $\thetak$ be its best $k$-term approximation, $\thetak_{q} \in \Theta$ the quantized version of
$\thetak$, for which we have
$$
\frac{1}{I^2}\| \thetak_{q} - \thetak \|^2_2 \le \frac{k}{m},
$$
and $\bar{\theta}^{(k)}_q$ the  element of
$\Gamma$ obtained by projecting $f^{(k)}_q = W \thetak_q$ onto $\cC$ and then transforming back into the basis $\Phi$: $\bar{\theta}^{(k)}_q = W^T \bar{f}^{(k)}_q$. Then, using (\ref{eq:projection}) and (\ref{eq:compressibility}), we get
\begin{eqnarray*}
\| \tf  - \bar{f}^{(k)}_q \|^2_2 &\le& \| \tf - f^{(k)}_q \|^2_2 \\
  &=& \| \ttheta - \thetak_q \|^2_2 \\
&\le& 2 \| \ttheta - \thetak \|^2_2 + 2 \| \thetak - \thetak_q \|^2_2 \\
&\le& I^2 \left( 2 C \rfm{\rho^2} k^{-2\alpha} + \frac{2k}{m} \right).
\end{eqnarray*}
Given each $1 \le k \le m$, let $\Gamma_k \subseteq \Gamma$ be the set
of all $\bar{\theta} \in \Gamma$, such that the corresponding $\theta
\in \Theta$ satisfies $\| \theta \|_0 \le k$. Then $|\Gamma_k| = {m
  \choose k} m^{k/2}$, so that $\log_2 |\Gamma_k| \le 2k \log_2 m$,
and therefore $\Gamma_k \in \cG_{N,p}$ whenever
$$
k \le k_*(N), \text{ where } k_*(N) \deq \frac{ N}{2c_4 \zeta_p^4 \log_2 m}.
$$
Then the first term on the right-hand
side of (\ref{eq:thm}) can be bounded by
\begin{eqnarray*}
\lefteqn{C_N \min_{1 \le k \le k_*(N)} R^*(\tf,\Gamma_k)} \\
&\le& O(N) \min_{1 \le k \le k_*(N)} \left[ \frac{1}{I^2} \| \ttheta - \bar{\theta}^{(k)}_q \|^2_2 + \frac{2 \pen(f^{(k)}_q)}{I}\right] \\
&\le& O(N) \min_{1 \le k \le k_*(N)} \left[ k^{-2\alpha} + \frac{k}{m} + \frac{k \log_2 m}{I}\right],
\end{eqnarray*}
where the constant obscured by the $O(\cdot)$ notation depends only on
$p$, $\rho$, $C$ and $c$.

\section{Auxiliary technical results}

\subsection{Proof of (\ref{eq:affinity})}
\label{app:affinity}

Given two Poisson intensity vectors $g, h \in \reals^N_+$, we have
\begin{eqnarray*}
\lefteqn{\int \sqrt{ p(y|g) p(y|h)} d\nu(y)} \\
&=& \prod^N_{i=1}\int \sqrt{ p(y_i|g_i) p(y_i|h_i)} d\nu_i(y_i) \\
&=& \prod^N_{i=1} \sum^\infty_{y_i = 0} \frac{(g_i h_i)^{y_i/2}}{y_i!} e^{-(g_i+ h_i)/2} \\
&=& \prod^N_{i=1} e^{-(g_i - 2(g_i h_i)^{1/2} + h_i)/2} \sum^\infty_{y_i = 0} \frac{(g_i h_i)^{y_i/2}}{y_i!} e^{-(g_i h_i)^{1/2}} \\
&=& \prod^N_{i=1}  e^{-\frac{1}{2}((g_i)^{1/2} - (h_i)^{1/2})^2} \underbrace{\int p\big(y_i\big|(g_i h_i)^{1/2}\big) d\nu_i(y_i)}_{=1} \\
&=& \prod^N_{i=1} e^{-\frac{1}{2}\left((g_i)^{1/2} - (h_i)^{1/2}\right)^2},
\end{eqnarray*}
where $\nu_i$ denotes the counting measure on the $i$th component of $y$. Taking logs, we obtain
$$
2 \log \frac{1}{\int \sqrt{p(y|g)p(y|h)}d\nu(y)} = \sum^N_{i=1} \left( (g_i)^{1/2} - (h_i)^{1/2}\right)^2.
$$
The quantity on the left-hand side is often used to measure divergence between probability distributions, and dates back to the work of Bhattacharyya \cite{Bha43} and Chernoff \cite{Che52}.

\subsection{Proof of (\ref{eq:li_barron})}
\label{app:li_barron}

For the sake of brevity, we will write $p_{\tf}(y)$ and $p_f(y)$ instead of $p(y|A\tf)$ and $p(y|Af)$. Also, define the {\em Hellinger affinity}
$$
\cA(\tf,f) \deq \int \sqrt{p_{\tf}(y) p_f(y)} d\nu(y).
$$
We then have
\begin{align*}
&2 \log \frac{1}{\cA(\tf,\hf)} = \\
& 2\log \left[\displaystyle\frac{\sqrt{p_{\hf}(y)/p_{\tf}(y)} e^{-\pen(\hf)}}{\cA(\tf,\hf)} \right] + \log \frac{p_{\tf}(y)}{p_{\hf}(y)} + 2\pen(\hf).
\end{align*}
In the first term on the right-hand side, the ratio is evaluated at $\hf$. Replacing this ratio by the sum of such ratios evaluated at every $f \in \Gamma$, we obtain the upper bound
$$
2 \log \sum_{f \in \Gamma} \left[\displaystyle\frac{\sqrt{p_f(y)/p_{\tf}(y)} e^{-\pen(f)}}{\cA(\tf,f)} \right] + \log \frac{p_{\tf}(y)}{p_{\hf}(y)} + 2\pen(\hf).
$$
Now we take expectation w.r.t.\ $p_{\tf}(y)$. Then, by Jensen's inequality,
\begin{align*}
& \expect_{\tf} \left\{ \log \sum_{f \in \Gamma} \left[\displaystyle\frac{\sqrt{p_f(y)/p_{\tf}(y)} e^{-\pen(f)}}{\cA(\tf,f)} \right] \right\} \\
& \qquad \le \log \sum_{f \in \Gamma} \left[ \frac{ e^{-\pen(f)}   } {\cA(\tf,f)} \underbrace{\expect_{\tf} \left\{ \sqrt{p_f(y)/p_{\tf}(y)} \right\}}_{=\cA(\tf,f)} \right] \\
& \qquad = \log \sum_{f \in \Gamma} e^{-\pen(f)} \le 0.
\end{align*}
By definition of $\hf$, we have
$$
\log \frac{p_{\tf}(y)}{p_{\hf}(y)} + 2\pen(\hf) \le \min_{f \in \Gamma} \left[\log \frac{p_{\tf}(y)}{p_{f}(y)} + 2\pen(f)\right].
$$
Thus,
\begin{align*}
& \expect_{\tf} \left[ \log \frac{p_{\tf}(y)}{p_{\hf}(y)} + 2\pen(\hf) \right] \\
& \qquad \le \min_{f \in \Gamma} \left[ \expect_{\tf} \log \frac{p_{\tf}(y)}{p_f(y)} + 2 \pen(f) \right] \\
& \qquad \equiv \min_{f \in \Gamma} \left[ \KL( p_{\tf} \| p_f ) + 2\pen(f)\right].
\end{align*}
Putting everything together, we get the bound
\begin{align*}
& 2\expect \log \frac{1}{\cA(\tf,f)} \\
& \qquad \le \min_{f \in \Gamma} \left[ \KL\Big( p(\cdot|A\tf) \Big\| p(\cdot|Af) \Big) + 2\pen(f) \right],
\end{align*}
as advertised.

\end{appendices}

\section*{Acknowledgments}

The authors would like to thank Emmanuel Cand\`es for several fruitful
discussions and the anonymous reviewers for very helpful suggestions.

\bibliographystyle{IEEEbib}
\bibliography{pcs_tsp_final.bbl}

\end{document}

%% file: myMacros.tex
\usepackage{amsmath,amssymb,latexsym}
\usepackage{stmaryrd}
\usepackage{amsfonts,bbm,booktabs}
\usepackage{times,lastpage,bm,xspace}
\usepackage{bibunits}
\usepackage[pdftex]{graphicx}

\usepackage{theorem}
\newtheorem{thm}{Theorem}
\newtheorem{remark}{Remark}
\usepackage[usenames]{color}
\definecolor{plum}  {rgb}{.4,0,.4}

\newcommand{\rfm}[1]{{#1}}
\newcommand{\rmw}[1]{{#1}}

 \newcommand{\xmath}[1]{{\ensuremath{#1}\xspace}}

 \def \Poisson{{\xmath{{\rm Poisson}}}}
 \def\tf{f^*}

\def \ttheta{\theta^*}
\def\tA{\widetilde{A}}

\def\pen{\mathop{\rm pen}}
\def\deq{\triangleq}
\def\thetak{\theta^{(k)}}

\def\hf{\widehat{f}}

\def\KL{\mathop{\rm KL}}

\def\jth{{\xmath{j^{{\rm th}}}}\xspace}

 \def\argmin{\mathop{\rm arg\,min}}
 \newcommand{\reals}{{\xmath{\mathbb{R}}}}
 
 \newcommand{\expect}{{\xmath{\mathbb{E}}}}
 \newcommand{\prob}{{\xmath{\mathbb{P}}}}

\newcommand{\squishlist}{
   \begin{list}{$\bullet$}
    { \setlength{\itemsep}{0pt}      \setlength{\parsep}{0pt}
      \setlength{\topsep}{0pt}       \setlength{\partopsep}{0pt}
      \setlength{\leftmargin}{1.5em} \setlength{\labelwidth}{1em}
      \setlength{\labelsep}{0.5em} } }

\newcommand{\squishlisttwo}{
   \begin{list}{$\bullet$}
    { \setlength{\itemsep}{0pt}    \setlength{\parsep}{0pt}
      \setlength{\topsep}{0pt}     \setlength{\partopsep}{0pt}
      \setlength{\leftmargin}{2em} \setlength{\labelwidth}{1.5em}
      \setlength{\labelsep}{0.5em} } }

\newcommand{\squishend}{
    \end{list}  }
\newtheorem{lemma}{Lemma}

\def\cF{{\cal F}}
\def\cG{{\cal G}}
\def\cC{{\cal C}}

\def\ave#1#2{\langle #1, #2 \rangle}
\def\1{{\bf 1}}